# Analyzing the Impact of Service Frequency and On-time Performance on Transit Ridership in Miami-Dade County

**Duanya Lyu**
Department of Civil & Coastal Engineering, University of Florida
1949 Stadium Rd, Gainesville, FL 32611, US
ORCID: 0000-0002-2468-8888

**Xiang Yan***
Department of Civil & Coastal Engineering, University of Florida
1949 Stadium Rd, Gainesville, FL 32611, US
ORCID: 0000-0002-8619-0065

* Corresponding author at:
Department of Civil & Coastal Engineering, University of Florida
1949 Stadium Rd, Gainesville, FL 32611, US
E-mail address: xiangyan@ufl.edu

**Acknowledgements**
We are grateful for the support received from Linda Morris at Miami-Dade Transit for this project. The authors thank Anran Zheng for gathering the stop arrival data from the Swiftly API.

**Funding**
This work was supported by the Center for Transit-Oriented Communities Tier-1 University Transportation Center (Grant No. 69A3552348337).

**Competing Interests Declarations**
The authors have no competing interests to declare that are relevant to the content of this article.

**Data Availability**
The General Transit Feed Specification Real Time data utilized in this research were sourced from the Swiftly, a mobility data provider, via its stop arrival API (https://swiftly.zendesk.com/hc/en-us/articles/360049238811-API-Guide-Basics). Due to licensing restrictions (https://www.goswift.ly/api-license ), this dataset is not publicly accessible. However, details on how to access the data and submit a purchase request are available on the provider's website.

**Author Contributions**
All authors contributed to the study conception and design. D.L. performed data preparation, analysis, result interpretation, prepared the initial manuscript draft. X.Y. performed analysis, result interpretation, review and editing, and secured funding. All authors commented on previous versions of the manuscript. All authors reviewed and approved the final manuscript.

**ABSTRACT**
This study investigates the impact of transit service attributes, focusing on service frequency and on-time performance (OTP), on bus ridership in Miami-Dade County. We obtained route-level ridership from automated passenger counter (APC) data and service performance metrics from the General Transit Feed Specification Real Time (GTFS-RT) data. The panel dataset allows us to effectively isolate the independent effects of OTP and frequency on bus ridership, contributing to a better understanding of how to improve service quality and support ridership. Our analysis examines both a pre-COVID period (January 2018-December 2019) and a post-COVID recovery period (July 2021-July 2023). Descriptive analysis reveals a steady ridership decline before the pandemic, followed by a sharp drop during the pandemic and a gradual recovery to pre-pandemic levels by late 2022. This recovery, however, varied by route and was accompanied by significant temporal fluctuations in service performance measures. To analyze these dynamics, we developed two-way fixed effects (2FE) models. Model outputs show that while frequency consistently influences ridership over time, OTP becomes an increasingly important determinant in the post-COVID recovery period. The observed significant interaction between frequency and OTP in the recovery period further suggests that improving reliability is particularly effective on frequent routes. Our study also highlights the need for targeted ridership-boosting strategies that differ by time of day, with a focus on punctuality during AM-peak and frequency during midday and PM-peak, to better support ridership growth and improve network performance.

## 1. INTRODUCTION

Public transit plays a vital role in providing mobility, reducing congestion, and lowering emissions (Lucas 2012). Effective transit planning requires understanding how transit ridership respond to both demand-side factors such as land use and demographics and supply-side attributes like frequency, reliability, and transit fares. Among supply-side attributes, *frequency* and *on-time performance (OTP)* are particularly influential (Taylor and Fink 2003). We have a general understanding of how the two shapes ridership: frequency influences rider attraction by reducing waiting times and congestion, and OTP is crucial for time-sensitive travelers who depend on reliable service (Kittelson & Associates, Inc. et al. 2013). However, the relative importance of these factors in shaping transit ridership and their potential interaction effects with various factors continue to elude us (Walker 2012). Additionally, the relationships between transit ridership and its influencing factors can vary widely across study areas and fluctuate over time, both longitudinally and by time of day, making it a complex empirical matter that warrants continued research efforts (Erhardt et al. 2022).

Understanding the relationship between supply-side attributes and transit ridership has become even more critical due to the disruptive impacts of COVID-19, which resulted in not only a sharp decline in transit demand but also travel patterns across time and space. The initial outbreak led to a 72.66% drop in ridership across 113 U.S. systems (Liu et al. 2020), followed by long-lasting shifts in who uses transit and how (Hu and Chen 2021; Liu et al. 2020). Preliminary studies suggest that the commuter patterns and time-of-day demand have changed (Liu et al. 2020). Riders also place greater emphasis on reliability (Hsieh 2023). In response, agencies reallocated services by adjusting frequency to match shifting demand, implemented operational strategies to improve reliability, and explored flexible options such as on-demand or microtransit services (Shortall et al. 2022). Despite the efforts, as of May 2024, most U.S. transit systems had not fully recovered (U.S. National Transit Database 2024). This highlights a critical need to study the impacts of these service changes on ridership to inform transit planning and operations in the post-pandemic era.

Due to the challenges of collecting fine-grained OTP data across space and time, empirical studies that examine the impact of OTP on transit ridership are limited in supply. The availability of General Transit Feed Specification Real Time (GTFS-RT) data, collected from onboard systems that track actual vehicle locations with timestamps, provides a solution to this empirical challenge. GTFS-RT data are often accessible through APIs provided by transit agencies, or providers who preprocess the data into aggregated outputs like stop arrival records, which can facilitate detailed service performance analysis (Genna 2024). Researchers have used GTFS-RT data to derive insightful metrics, such as observed headways, on-time rates, and bus delays, based on actual service performance rather than scheduled times. (Aemmer et al. 2022).

This study leverages a multi-year panel dataset to examine how transit service attributes, specifically frequency and OTP, shape ridership patterns at the route level. We focus on Miami-Dade Transit's bus service, with analysis spanning two distinct periods: a pre-COVID period between January 2018 and December 2019 and a post-COVID recovery period between July 2021 and July 2023. By applying two-way fixed effects (2FE) models to examine high-resolution GTFS-RT and APC data, we examine the evolving service performance-ridership relationship. The objectives are twofold: 1) to empirically analyze route ridership and service changes during the pre-COVID and recovery periods, and 2) to model and compare how service frequency and OTP influence ridership across both periods. These findings contribute to a broader understanding of how service quality influences ridership and offer practical guidance for transit planning and operations in both stable and transitional service environments.

## 2. LITERATURE REVIEW

### 2.1 Factors Influencing Bus Ridership

Factors influencing ridership can be classified into the demand-side and the supply-side (Taylor and Fink 2003). *Demand-side* factors include socio-economic indicators such as income, employment rate, car ownership, and competitive or complementary modes (Hall et al. 2018). *Supply-side* factors include service quantity and service quality (Soza-Parra et al. 2021). As for quantity, larger coverage and frequency attract more riders, but limited resources often necessitate a coverage-frequency trade-off (Berrebi et al. 2021; Kittelson & Associates, Inc. et al. 2013). Service quality, encompassing factors like OTP and headway adherence, is equally important, as it reflects the reliability of the transit options and determines the *realizable* accessibility (Liu et al. 2024). Advancements in positioning and communication technology have increased attention to these metrics in recent studies (Aemmer et al. 2022).

The effects of these operational factors are multifaceted and context-dependent. Rider perceptions and requirements of them are usually time-sensitive, meaning the importance of operational factors can vary by time of day. For instance, commuters often value punctuality more due to work constraints (Bowman and Turnquist 1981). Moreover, there may be diminishing marginal returns as service quality or quantity increases (Berrebi et al. 2021). Additional frequency on an already well-served route may result in only modest increases in ridership. The impact of one variable (such as frequency) may also be influenced by the levels of another (such as OTP), emphasizing the interdependent nature of supply-side factors (Walker 2012).

Researchers have been using direct demand modeling techniques, such as four-step models, to analyze the relationship between demand and supply systematically (Boyle 2006). There are also regression models that describe the relationships given cross-sectional observations (Lucas Albuquerque-Oliveira et al. 2024). However, supply-side factors are not independent of demand-side factors, as transit agencies adjust service levels based on ridership demand, which introduces endogeneity (Kerkman et al. 2015). Therefore, models may be biased due to the omitting variable problem and the supply-demand endogeneity, especially when applied to period data (Berrebi et al. 2021; Kerkman et al. 2015). To address these issues, researchers have adopted panel data analysis techniques, which allow better representation of observation periods and units. Panel data models are also more effective in addressing endogeneity (Yaffee 2003). Studies have shown significant differences between estimations from cross-sectional models and panel data models, highlighting advantages of the approach (Berrebi et al. 2021; Kerkman et al. 2015).

### 2.2 Bus Ridership in the COVID-19 Era

A number of studies have examined the impact of the COVID-19 pandemic on transit demand. Demographically, the dynamics varied across population groups, for example, the impact on commuters (Shimamoto and Kusubaru 2023) and high-income neighborhoods are smaller (Wilbur et al. 2023). The pandemic also shifted the time-of-day demand patterns. For example, peak-hour ridership saw a larger decline during the pandemic, making the time-of-day distribution less peaky, while weekend ridership curves became peakier and more similar to weekday curves (Liu et al. 2020). Regarding peak periods, the morning (AM) and afternoon (PM) peaks also behaved differently: the AM-peak saw a larger mid-pandemic decline (Wilbur et al. 2023), while the PM-peak shifted to start earlier and last longer (Liu et al. 2024). Furthermore, survey-based studies indicated a stronger emphasis on service quality among transit riders (Hsieh 2023; Kapatsila et al. 2023). These observed demand-side changes in ridership patterns and rider preferences may have lasting impacts considering the shifts in transit rider populations prompted by factors like permanent remote work and owning vehicles (Hu and Chen 2021; Shimamoto and Kusubaru 2023).

Researchers have also examined transit agencies' operational responses to transit demand changes. Shortall et al. reviewed COVID-era operational changes, classifying them into "avoid," "modal shift," and "quality improvement," highlighting limited assessments of their social impacts in literature (Shortall et al. 2022). Dasmalchi and Taylor found that the dramatic shifts in transit demand across Boston, Houston, and Los Angeles during the pandemic has caused disparities in service supply, but the mismatch was partly mitigated by the service changes in late 2020 (Dasmalchi and Taylor 2022). Wilbur et al. analyzed operational changes and ridership data, finding the initial ridership decline largely uncorrelated with supply (vehicle runs) (Wilbur et al. 2023). Gkiotsalitis et al. reviewed traditional tactical and operational planning models for the post-pandemic era, highlighting the need to adjust planning measures (e.g., previous focus on efficiency and passenger numbers often led to overcrowding). Moreover, the changes in ridership patterns and rider preferences discussed above call for more flexible service (Gkiotsalitis and Cats 2021). For example, the flattening of peak hour demand may make some transit agencies consider reducing service frequencies during peak hours. However, such adjustments could diminish the appeal of public transit for many and result in further decline in ridership. For instance, Redelmeier et al. constructed a longitudinal panel with a COVID-19 dummy and found preliminary evidence of an increased elasticity of transit ridership to service frequency after the pandemic (Redelmeier and El-Geneidy 2024).

In sum, existing studies have focused more on the associations between ridership and demand-side factors such as areal or rider characteristics, with limited attention paid to supply-side factors such as service frequency and OTP. Ridership patterns and the underlying relationships between service performance and ridership have been changing over time, particularly when considering periods before and after the pandemic. Due to the lack of systematic studies in this regard, much is still unknown regarding how transit ridership would respond to and feedback the service and operational changes made by various agencies in the post-pandemic era. With more stable patterns and additional data available in the post-pandemic period, there is now an opportunity to generate novel insights about ridership dynamics that can have lasting scientific value.

## 3. STUDY AREA AND DATA

### 3.1 Study Area

Miami-Dade County, located in the southeastern part of Florida, is home to over 2.7 million residents, making it the most populous county in the state. The county boasts a diverse demographic composition and serves as a major hub for finance, commerce, culture, arts, and international trade (U.S. Census Bureau 2022). Miami-Dade County is home to the largest transit system in Florida and the fifteenth-largest in the U.S. (County of Miami-Dade Transportation & Public Work 2021). Miami-Dade Transit (MDT) operates the public transportation system within the county, which includes metro rail, bus, and mover services. Metrobus, the bus service component, operates around 100 routes and forms the backbone of the county's transit network, accounting for over half of total ridership, providing extensive coverage across urban and suburban areas. This network not only supports daily commutes but also plays a vital role in connecting residents to essential services and amenities throughout the county. Although severely impacted by the pandemic, Metrobus ridership rebounded strongly, reaching 4.55 million in June 2023, even 19.9% above its June 2019 levels (Better Streets Miami Beach 2023). This substantial ridership and robust recovery make the bus service an ideal case for analyzing ridership patterns and informing post-pandemic transit strategies.

### 3.2 Data

We obtained monthly route ridership data from the Automatic Passenger Counting (APC) dataset provided by the agency. This dataset offers average weekday boardings at the route level for each month. For service performance, we extracted GTFS-RT data using the Swiftly API (Aemmer et al. 2022). Swiftly preprocesses the raw GTFS-RT feed into a stop-level arrivals format. **Table 1** lists the specific data fields requested. Based on the data, we derived key service performance metrics, including headways and on-time rates, as detailed in the following sections (See *Section 4.1*). According to Swiftly, it applies quality control procedures to address recording errors and inconsistencies in the raw feed (Warade 2025). To further assess data completeness, we randomly selected ten days in 2022 and compared against agency schedule data and found that 93% of scheduled trips and 84% of scheduled stop arrivals were captured.

**Tab.1 Example GTFS-RT Data Obtained from Swiftly**

| | *Data Entry 1* | *Data Entry 2* | *Data Entry 3* |
|---|---|---|---|
| *trip_id* | 4055703 | 4055703 | 4055694 |
| *route_id* | 19324 | 19324 | 19382 |
| *direction_id* | 1 | 1 | 0 |
| *route_short_name* | 1 | 1 | 119 |
| *stop_id* | 3674 | 13 | 13 |
| *stop_name* | SW 211 St @ SW 112 Av | SW 211 St @ SW 107 Av | SW 211 St @ SW 107 Av |
| *vehicle_id* | 22136 | 22136 | 19244 |
| *gtfs_stop_seq* | 17 | 18 | 18 |
| *start_time* | 11:15 | 11:15 | 17:15 |
| *scheduled_date* | 6/1/2023 | 6/1/2023 | 6/1/2023 |
| *scheduled_time* | 11:33:03 | 11:35:00 | 17:35:00 |
| *actual_date* | 6/1/2023 | 6/1/2023 | 6/1/2023 |
| *actual_time* | 11:35:08 | 11:36:06 | 17:40:21 |

We also include additional variables capturing land use and sociodemographic conditions, as prior studies have shown they significantly influence ridership patterns (Taylor and Fink 2003). Land use data were sourced from parcel-level land-use datasets from the Florida Geo Database Library (FGDL), covering each year within the study period (University of Florida GeoPlan Center 2024). The dataset includes detailed information on various land use types, allowing us to capture the proportion of land dedicated to different uses within route catchment areas. Sociodemographic data were obtained from the U.S. Census Bureau's American Community Survey (ACS) 5-year estimates for each year (U.S. Census Bureau 2022). These data were collected at the census tract level and aggregated to route catchment areas (See *Section 4.3*). Including the two sets of control variables helps us control for the potential changes in urban conditions over time.

## 4. METHOD

### 4.1 Operationalization of Key Performance Measures

*(1) Service Frequency*

Frequency, a crucial measurement for service supply, is related to bus service accessibility (Lucas Albuquerque-Oliveira et al. 2024) and directly influences riders' decisions to use transit (Diab et al. 2021). Here we follow common practice to operationalize frequency with averaged headway of a route (Redelmeier and El-Geneidy 2024). Specifically, we first calculated stop-level headways as the difference in actual arrival times between consecutive vehicles at each stop. To match the monthly averaged weekday ridership, these headways at stops were then averaged across stops within a trip of the route, across the trips within each day, and finally across the month to derive the headway measurement $hw_{it}$ for route $i$ in month $t$:

$$hw_{it} = avg_{d \in D_{it}} \left( avg_{tr \in TR_{d,it}} \left( avg_{s \in S_{tr,d,it}} \left( ts_{s,tr,d,it} - ts_{s,(tr-1),d,it} \right) \right) \right) \quad (1)$$

where $\left( ts_{s,tr,d,it} - ts_{s,(tr-1),d,it} \right)$ is the actual arrival time interval (i.e. headway) at stop $s$ between trip $tr$ and its preceding trip $(tr - 1)$ for route $i$, $S_{tr,d,it}$ is the set of all stops visited during trip $tr$, $TR_{d,it}$ is the set of all trips conducted within a specific day $d$, and $D_{it}$ is the set of all days in month $t$.

*(2) On-time Performance*

OTP reflects service reliability and is an essential consideration in passenger satisfaction (Danaher 2003). The OTP of a route is influenced by many factors including route type, stop location, scheduling, and time-of-day traffic conditions (Walker 2012). In this study, we measure route OTP with the *average on-time rate* across its stops. In calculating, we start with the ratio of on-time arrivals across all stops for each trip; these ratios were then averaged across trips within a day, and subsequently averaged across days in a month to yield the route-level monthly averaged on-time rate measurement $ot_{it}$:

$$ot_{it} = avg_{d \in D_{it}} \left( avg_{tr \in TR_{d,it}} \left( \frac{\sum_{s \in S_{tr,d,it}} \delta_{s,tr,d,it}}{\left| S_{tr,d,it} \right|} \right) \right) \quad (2)$$

where $\delta_{s,tr,d,it}$ is a binary indicator (1 if on-time, 0 if not) for whether an arrival was on-time. In this study, "on-time" is defined as arriving within two minutes early to five minutes late (Arhin and Noel 2013; TransitCenter 2018). While another common definition is a one-minute early threshold, the broader two-minute window aligns with local agency practice and better reflects operating conditions. We acknowledge that an overall measurement of average on-time rate can mask extremes at stops. We used the metric because it gives a basic level of route OTP, aligns with agency KPIs and yields interpretable elasticities.

Relying solely on the average on-time rate can obscure the potential temporal instability in OTP: a route may exhibit a steady average on-time rate while experiencing substantial trip-to-trip fluctuations. To capture this, we complement the mean on-time rate with a measure of on-time rate variability. Here, we first compute the trip-level on-time rates as before. Within each day, we then calculate the standard deviation of the trip-level on-time rates across all trips operated in the day. Finally, we aggregate across the month by averaging these daily standard deviations, yielding a monthly measurement of route OTP consistency:

$$otv_{v_{it}} = avg_{d \in D_{it}} \left( stddev_{tr \in TR_{d,it}} \left( \frac{\sum_{s \in S_{tr,d,it}} \delta_{s,tr,d,it}}{\left| S_{tr,d,it} \right|} \right) \right) \tag{3}$$

### 4.2 Modeling Approach

To examine the relationship between service performance and bus ridership, we adopt a fixed-effects (FE) modeling framework. FE models leverage panel data to control for unobserved heterogeneity that is constant across either units or time, thereby reducing omitted variable bias and yielding more reliable estimates than cross-sectional approaches (Yaffee 2003). When including both unit- and time-fixed effects with a two-way fixed-effects (2FE) specification, the model is formally expressed as:

$$y_{it} = \boldsymbol{X}_{it}\boldsymbol{\beta} + u_i + v_t + e_{it}, \tag{4}$$

where $y_{it}$ is outcome variable (e.g., ridership) for route $i$ at month $t$, $\boldsymbol{X}_{it}$ is the vector of explanatory variables, $u_i$ and $v_t$ represents the unit- and time-fixed effects, respectively, and $e_{it}$ is the error term. A common estimator for the 2FE model is the within estimator, which excludes unit-specific time average ($\overline{x_{\iota t}}$) and applies pooled Ordinary Least Squares (OLS) for estimation (Croissant and Millo 2008):

$$y_{it}' = y_{it} - \overline{y_t} = (x_{it} - \overline{x_{\iota t}})\boldsymbol{\beta} + (e_{it} - \overline{e_\iota}) \tag{5}$$

The 2FE approach is particularly suitable for ridership modeling, where both route-level differences and external temporal shocks play important roles (Berrebi et al. 2021; Hall et al. 2018; Kerkman et al. 2015). Including route-fixed effects (i.e. unit-fixed effects) controls for time-invariant differences across routes, such as service region, built environment, or broader demographic context; including time-fixed effects captures system-wide shocks, such as macroeconomic conditions, seasonal patterns, or COVID-related disruptions (Hall et al. 2018). The 2FE structure enables us to more precisely isolate the effects of key service performance measures net of unobserved spatial and temporal confounders.

In this study, we estimate the 2FE models with the plm() package in R (Croissant and Millo 2008). We compared this 2FE specification against pooled OLS, one-way FE and random effects (RE) models. F-test is used to assess the significance of route- and time-fixed effects, and Hausman test is used to determine whether RE estimates are consistent and preferable (Croissant and Millo 2008).

### 4.3 Model Specification

In this study, we adopt a log-log specification of 2FE model. In a log-log specification, the coefficients represent elasticities, indicating the percentage change in ridership from a 1% change in an explanatory variable, holding others constant. This approach has been widely used to understand the relative responsiveness of ridership to changes in various factors (Berrebi et al. 2021; Diab et al. 2021). We specify the following log-log 2FE model:

$$log(\text{Risdership}_{it}) = \beta_1 log(HW_{it}) + \beta_2 log(OTR_{it}) + \beta_3 log(HW_{it}) \times log(OTR_{it})$$
$$+\boldsymbol{\beta}_{ctrl} \log(\boldsymbol{X}_{ctrl_{it}}) + u_i + v_t + e_{it}, \tag{6}$$

where $Ridership_{it}$ represents the dependent variable: average weekday ridership for route $i$ at month $t$; the two service performance measurements of interest are headway $HW_{it}$ and on-time rate $OTR_{it}$ with their interaction term; the control variable set is denoted by $\boldsymbol{X}_{ctrl_{it}}$; $u_i$ and $v_t$ are the route- and time-fixed effects, respectively; $e_{it}$ is the error term.

***Interaction and Lagged Variables***

We included an interaction term between headway and on-time rate to capture the potential moderating effect of one service performance measure on the influence of the other (Preacher and Rucker 2003). This is relevant as riders' priorities may differ by service frequency: those on low-frequency routes often place greater importance on punctuality (Walker 2012). Under this specification, a significant interaction term indicates that the effect of OTP ($\beta_2 + \beta_3 \times log(HW)$) varies depending on headway.

We also lagged the three service performance measures by using a two-month moving average of the two preceding months. This is to account for the delayed influence of service changes, reflects the notion that ridership adjustments often take time to materialize (Chen et al. 2011; Yu et al. 2024). Averaging across two months provides a more stable representation of recent service conditions than a single-month lag, which may be overly sensitive to short-term fluctuations (e.g., weather, construction). At the same time, it avoids excessive smoothing that could occur with a longer window, which might dampen the responsiveness of ridership to recent changes. and avoids the over-smoothing of longer windows that can dampen responsiveness.

***Control Variables***

To account for the contextual factors including ridership beyond service performance, we include a set of control variables considering service configuration, sociodemographic characteristics, and land use composition.

For service configuration, we included a measure of trip length to account for route structure in the study periods. We gathered the GTFS Static (GTFS-ST) data published by the agency. To correspond to weekday route ridership data, we extracted the regular weekday service from the GTFS-ST data by identifying a typical weekday trip and used the trip length. From the data, we also identified the stops in the weekday trips and buffered 0.25-mile around them to define the route catchment areas, which were then utilized to calculate route-level sociodemographic characteristics and land use composition controls.

Candidate sociodemographic variables in this study included population, working-age population in the labor force, population with a bachelor's degree, population identified as white race only, household with no vehicle, median household income. Candidate land use categories included residential, retail, recreation, public and institutional land. Route-level sociodemographic and land use characteristics were constructed using the ACS data and parcel land use data mentioned in *Section 3.2* and aggregated based on the intersection area weighted by service route catchment areas mentioned in the previous paragraph. To address collinearity, we excluded variables with pairwise correlations above 0.6 or variance inflation factors (VIF) greater than 10. The final specification retained median household income vehicles, labor force share, and share of white-alone race population for sociodemographic controls, and the percentage of residential, retail, and recreational land within each route's catchment area for land use controls.

Unlike service performance measures, control variables inputs are not lagged, because both land use and sociodemographic data were at an annual frequency. Given that GTFS-ST data were updated biannually, we incorporated potential route alignment changes by updating the calculations biannually based on the GTFS-ST route geometry.

***Model Estimation***

We split the data and estimate separate models for two periods: a pre-COVID period between January 2018 and December 2019 and a post-COVID recovery period between July 2021 and July 2023. The early pandemic period (2020 to mid-2021) was not analyzed because it experienced significant and temporary disruptions in both service provision and travel behavior, which would confound analysis of longer-term trends and service-ridership relationships. We separate the two periods for two main considerations. First, service patterns and network structure changed during and after the pandemic, including new route launches and route consolidations (CBS Miami 2020b; CBS Miami 2020a). More importantly, prior research indicates that ridership dynamics differ between the pre- and post-COVID periods, particularly regarding sensitivity to service performance (Hsieh 2023; Redelmeier and El-Geneidy 2024). Separating the two periods allows us to capture these differences meaningfully and test the robustness of findings across distinct operational contexts.

Furthermore, recognizing that rider characteristics and priorities differ by time-of-day, we estimate separate models for different time-of-day periods to capture the potential temporal heterogeneity in the relationship between ridership and service performance. We consider four time-of-day periods: *AM-peak* (6:00–8:59), *Midday* (9:00–15:59), *PM-peak* (16:00–18:59), and *Early Evening* (19:00–21:59). Because the coefficients in this log-log specification are elasticities and each regression includes route- and month-fixed effects, this design enables direct comparison of elasticities across time-of-day periods. To derive ridership for each period, we used a supplemental trip-level APC dataset containing annually averaged, time-stamped stop-level boardings. For each route and time-of-day period, we calculated the share of daily boardings and applied these proportions to total daily ridership to estimate time-of-day-specific values. We also calculated service performance metrics using only the trips operating within each period, ensuring the measures accurately reflect the time-of-day-specific conditions.

# 5. RESULTS

## 5.1 Descriptive Analysis on Key Variables

### *5.1.1 Summary Statistics and Variable Distributions*

**Table 2** presents some descriptive statistics for ridership and the two key service performance measurements for the pre-COVID period and the recovery period. *Ridership* was the highest for Midday, reflecting its longer duration (9:00 AM–3:59 PM). AM-peak ridership slightly exceeded PM-peak, both of which were higher than Early evening. This pattern also held for ridership standard deviations. After COVID, ridership declined across all periods, with the largest drop observed for midday (to 85% of the pre-COVID level) and the smallest for AM-peak (87%). The standard deviation of ridership also decreased, suggesting smaller variation. *Headways* were shortest during the AM-peak, followed by PM-peak, Midday, and Early evening. Post-COVID, despite reduced ridership, headways shortened across all periods, indicating increased frequency. The variability in headway also declined, except during Midday, implying more consistent service frequency across routes. *On-time rate* was highest in the AM-peak, followed by Midday, Early evening, and PM-peak. Post-COVID, on-time rate improved across all periods, suggesting better service OTP. The standard deviation of on-time rate declined, except during PM-peak, indicating larger OTP variation for that period.

**Tab.2 Descriptive Statistics of Key Variables**

| Period | Time-of-Day | Ridership | Headway | On-time Rate |
|---|---|---|---|---|
| **Pre-COVID Period** | AM-peak | 677.4 (435.51) | 30.09 (13.32) | 0.66 (0.11) |
| | Midday | 1327.28 (1021.91) | 35.48 (12.68) | 0.58 (0.12) |
| | PM-peak | 603.92 (406.12) | 32.92 (14.22) | 0.48 (0.14) |
| | Early evening | 197.60 (193.3) | 41.11 (13.86) | 0.53 (0.17) |
| | All day | 701.55 (723.05) | 34.83 (14.12) | 0.56 (0.15) |
| **Recovery Period** | AM-peak | 589.39 (426.84) | 28.61 (12.16) | 0.66 (0.10) |
| | Midday | 1128.22 (930.01) | 33.57 (13.48) | 0.60 (0.11) |
| | PM-peak | 520.57 (383.07) | 30.64 (13.34) | 0.50 (0.15) |
| | Early evening | 170.97 (175.27) | 40.07 (13.67) | 0.56 (0.16) |
| | All day | 602.29 (650.63) | 33.15 (13.85) | 0.58 (0.15) |

* Values are presented as mean followed by standard deviation in parentheses

**Figure 1** shows the distributions of the variables. The y-axis represents density, meaning the proportion of observations per unit interval. For *ridership*, Midday period shows a wider spread, while the Early evening is concentrated at lower values, with AM-peak and PM-peak in between. The distribution patterns for ridership remain broadly similar across the pre-COVID period and recovery periods, though the recovery period shows more concentration at lower values, particularly during early evening. *Headways* are shorter during AM-peak and PM-peak, especially in the AM-peak, while the Midday period tends to have larger but more concentrated headways around the mean and Early evening observations show more observations of longer headways. Compared to the pre-COVID period, Midday headway in the recovery period showed wider distribution, while AM-peak and PM-peak show tighter clustering around shorter headways. For *on-time rate*, AM-peak consistently performs better, with more observations achieving high values, while Midday and Early evening periods show greater spread toward lower values. In the recovery period, AM-peak, Midday, and Early evening periods all show slight improvements with more observations at higher on-time rates, but PM-peak shows a wider spread with lower-performing observations.

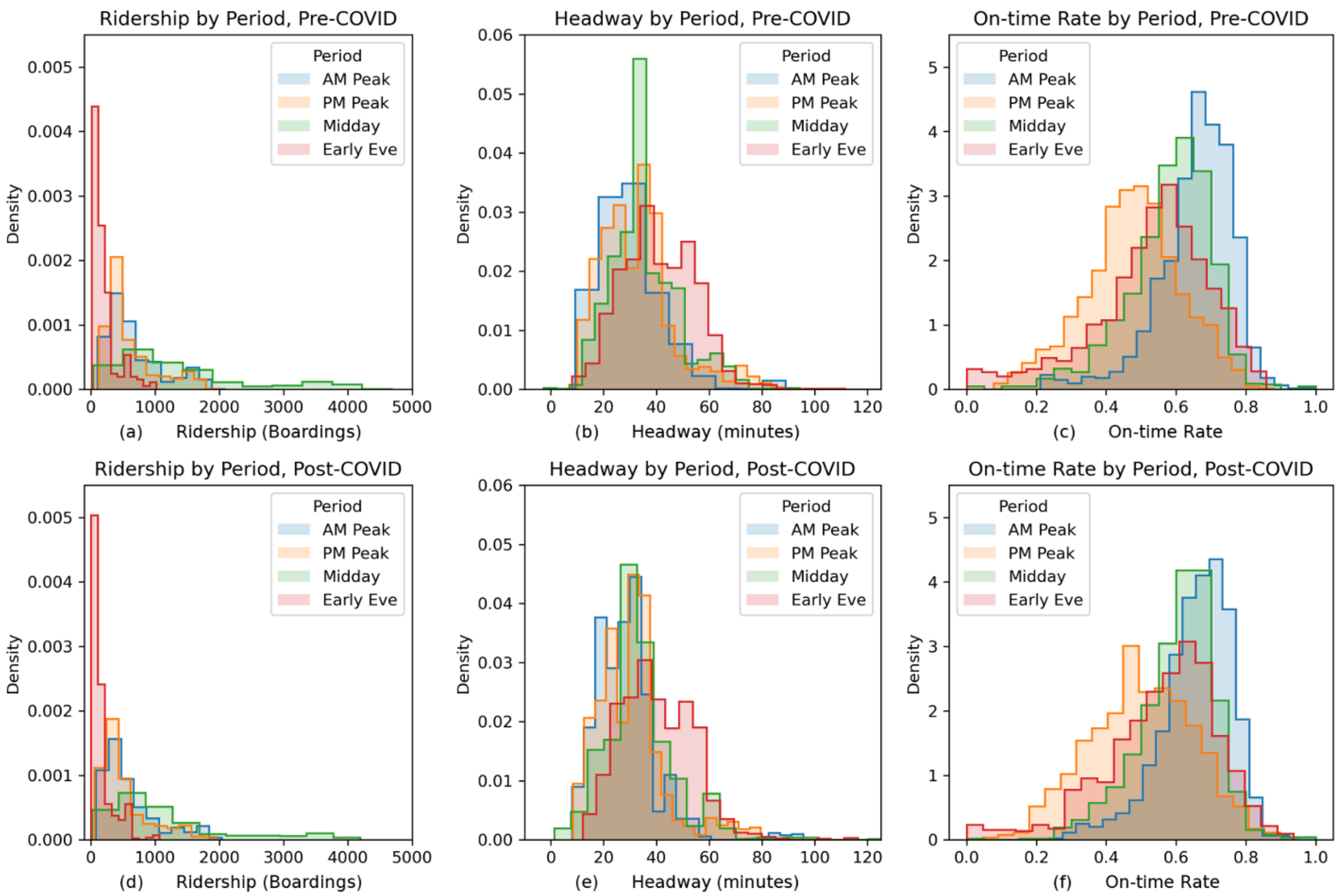

**Fig.1 Distribution of Key Variables**

***5.1.2 Temporal Trends***

**Figure 2** shows temporal trends of ridership and the two performance measures. Thick curves represent averages across routes, while thin lines represent individual routes. We can see that *ridership* in Miami-Dade County was already declining before the pandemic, mirroring national trends driven by demand-side factors such as ride-hailing, low gas prices, telework, and increased car ownership (Erhardt et al. 2022). COVID-19 accelerated this decline, particularly during the Midday; though, steady recovery followed, possibly due to its core base of blue-collar riders (Espinoza 2020). Ridership trends across all time-of-day periods follow a similar pattern. *Headway* curves display fairly consistent patterns across all four time-of-day periods for both pre-COVID and recovery periods. The curves are slightly more compressed during the recovery period. Early in the recovery period, some routes saw increased frequency, possibly to reduce crowding (Espinoza 2020); though, the additions were later reduced to previous level around early 2022. *On-time rates* curves show large variation with evident seasonal patterns, with slight dips during winter and peaks in the summer months. We also see distinct patterns across the four time-of-day periods. AM-peak and Midday show higher punctuality and smaller fluctuations compared with Early evening and PM-peak. Post-COVID, AM-peak and Midday and Early evening show better punctuality with smaller variation, indicating successful operational adjustments and potentially less congestion (Lee et al. 2020). PM-peak, however, retains its low performance.

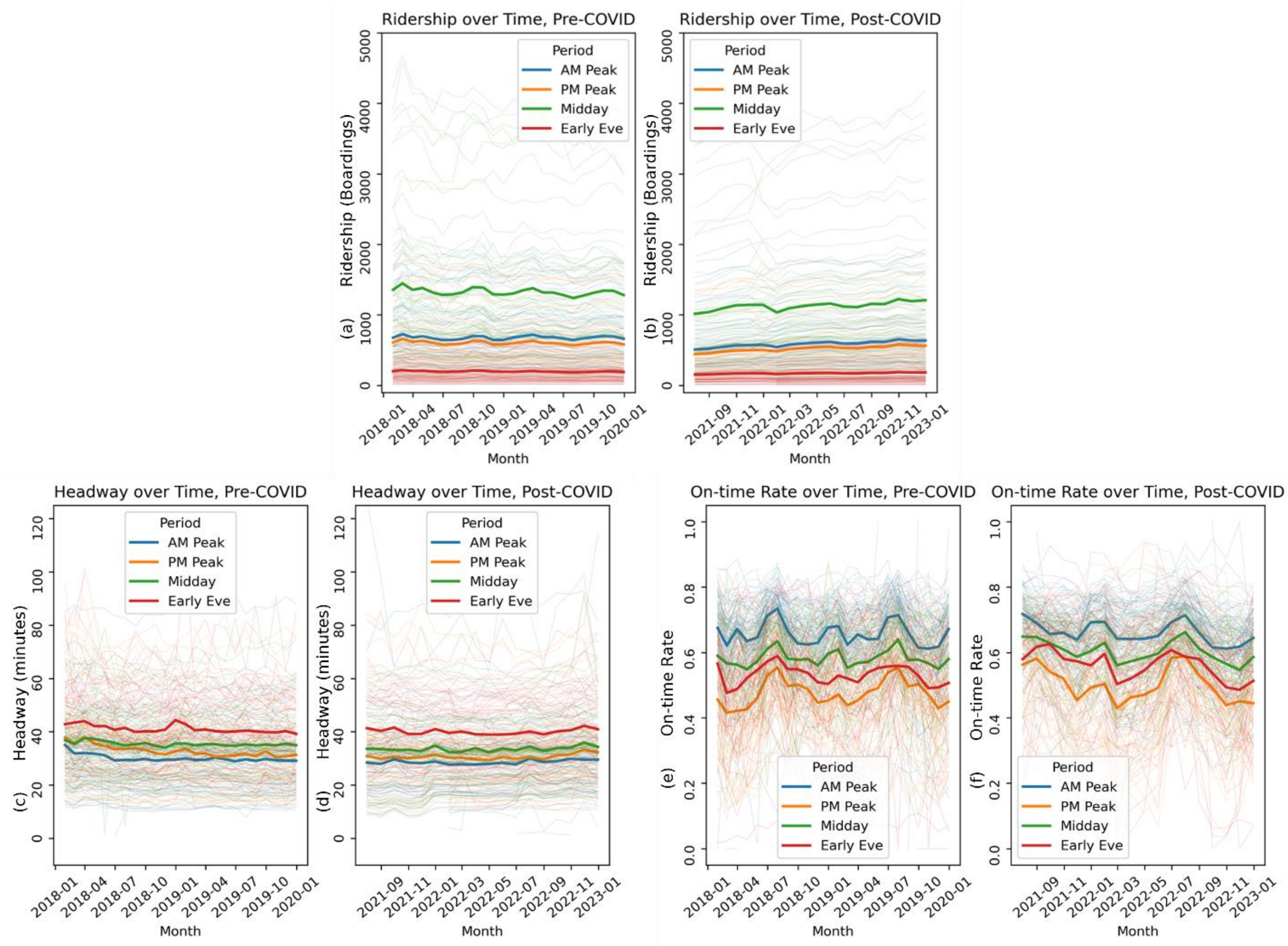

**Fig.2 Temporal variations of key variables in the pre-COVID and recovery periods**

### *5.1.3 Spatial Patterns*

**Figure 3** shows the spatial distributions of ridership and the two key variables of interest in the pre-COVID and recovery periods. *Ridership* was particularly high along major transit corridors and densely populated urban centers before COVID, reflecting areas with significant commuting demand. In the recovery period, ridership declined overall, with a more uneven spatial distribution. Those routes serving suburban or peripheral areas saw particularly significant decreases, while core urban routes showed larger ridership, suggesting a prioritization of transit use in these areas even during the recovery period. This likely reflects shifts in travel behavior due to remote work, population movement, or altered commuting habits in the study area (Department of Transportation and Public Works, Miami-Dade County 2023). *Headway* also shows notable shifts. Before COVID, frequent service was offered on major routes to accommodate high ridership volumes. During the recovery period, many routes had increased headways, particularly in areas with declining ridership. The map further highlights a shift towards concentrating service at central areas, although it may come at the expense of reducing transit accessibility for some areas. *On-time rate* was uneven across the network. Routes with higher on-time rates were generally located in suburban areas, where traffic congestion is less severe. Conversely, central urban routes exhibited lower on-time rate, reflecting traffic-related delays. During the recovery period, efforts to improve OTP appeared to benefit suburban and peripheral routes more.

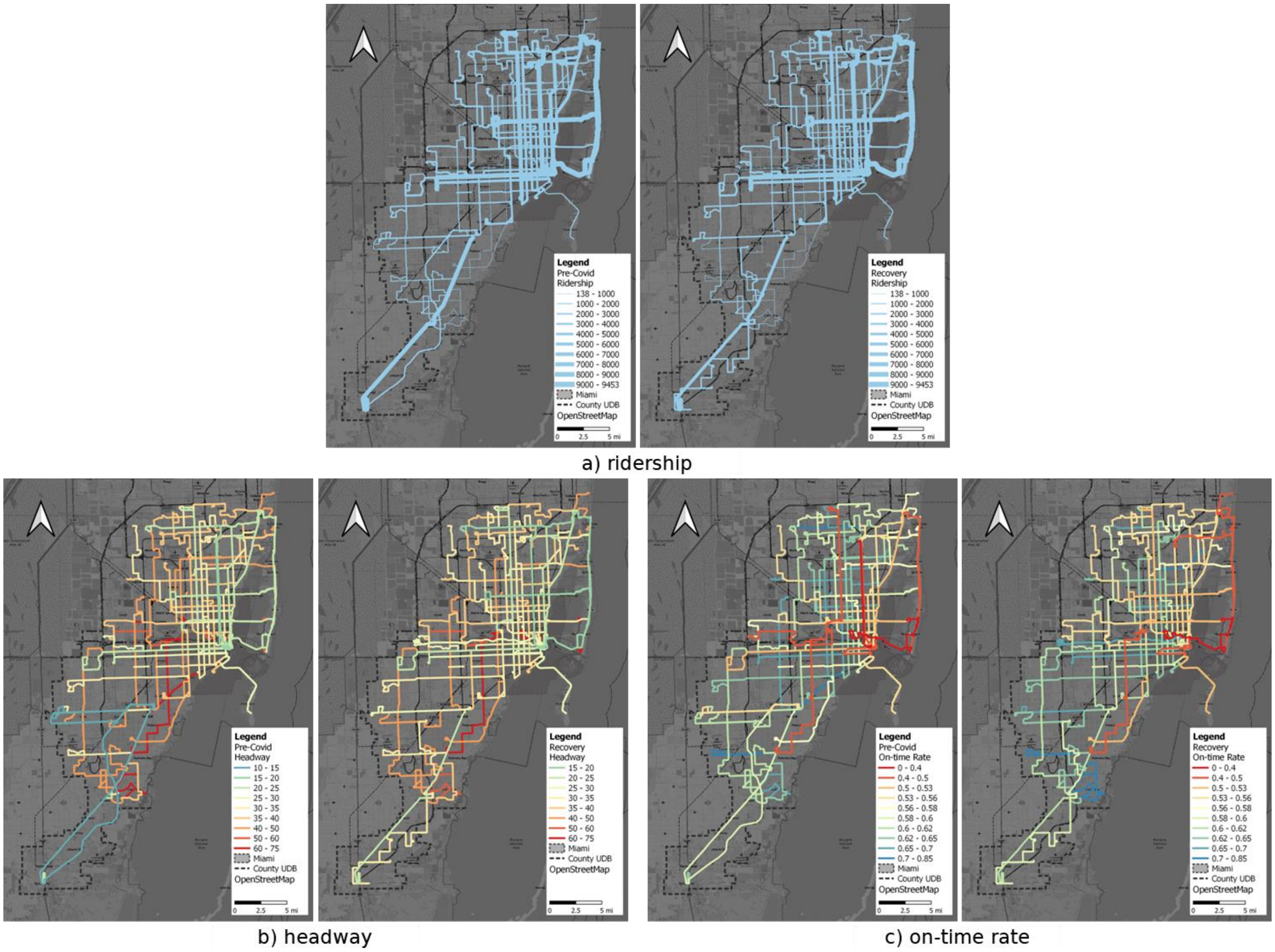

**Fig.3 Spatial distributions of key variables across pre-COVID and recovery periods**

## 5.2 Model Outputs

To validate the choice of the 2FE approach, we compared multiple model specifications, including one-way FE (by route or by time), one-way random effects (RE), and two-way RE models. We first conducted Hausman tests, and the results showed significant differences between fixed and pooled/random effects models, indicating correlation between regressors and unobserved effects. We then compared among the FE specifications with F-tests and found significant joint significance of the included two-way effects, indicating that accounting for both route and time effects better captures unobserved heterogeneity. To ensure model robustness, we tested for serial correlation and heteroscedasticity. The Pesaran CD test indicated significant cross-sectional dependence, suggesting unobserved factors correlate across entities; the Breusch-Godfrey/Wooldridge test revealed significant serial correlation, implying correlation in error terms over time. Given both types of correlation, we employed Driscoll-Kraay standard errors for robust estimates (Wooldridge 2010).

**Table 3** presents the estimated coefficients from the 2FE models. Each cell displays the coefficient estimate with its significance level, followed by the *t*-statistic and *p*-value in parentheses.

**Tab.3 Model Estimation Results**

| VARIABLE | AM Pre-COVID | AM Recovery | PM Pre-COVID | PM Recovery | MIDDAY Pre-COVID | MIDDAY Recovery | EARLY EVENING Pre-COVID | EARLY EVENING Recovery |
|---|---|---|---|---|---|---|---|---|
| *HW* | **-0.100 ***** | **-0.202 *** | **-0.110 **** | **-0.274 ***** | **-0.142 *** | **-0.315 ***** | **-0.192 **** | 0.016 |
| | **(-3.37, <0.001)** | **(-2.06, 0.040)** | **(-2.73, 0.0063)** | **(-3.82, <0.001)** | **(-2.16, 0.0306)** | **(-5.12, <0.001)** | **(-3.05, 0.0024)** | (0.27, 0.785) |
| *OTR* | 0.097 | **0.877 *** | -0.007 | **0.417 *** | -0.006 | **0.584 *** | **0.396 *** | -0.012 |
| | (0.46, 0.649) | **(2.20, 0.0282)** | (-0.07, 0.946) | **(2.16, 0.0312)** | (-0.02, 0.985) | **(2.01, 0.0448)** | **(2.23, 0.0257)** | (-0.05, 0.960) |
| *OTR×HW* | -0.026 | **-0.266 *** | 0.013 | **-0.142 **** | 0.007 | **-0.191 **** | **-0.112 *** | 0.005 |
| | (-0.40, 0.692) | **(-2.22, 0.0270)** | (0.47, 0.637) | **(-2.64, 0.0085)** | (0.08, 0.935) | **(-2.75, 0.0062)** | **(-2.40, 0.0166)** | (0.07, 0.944) |
| *OTV* | 0.045 | 0.024 | 0.020 | 0.015 | -0.001 | **0.104 ·** | -0.017 | **-0.061 *** |
| | (1.33, 0.184) | (0.37, 0.710) | (0.84, 0.400) | (0.49, 0.622) | (-0.11, 0.913) | **(1.81, 0.071)** | (-1.05, 0.294) | **(-2.30, 0.021)** |
| *WHITE_PCT* | -0.214 | **-0.204 **** | -0.090 | **-0.336 ***** | -0.038 | -0.082 | **0.356 ·** | **-0.175 ***** |
| | (-1.54, 0.124) | **(-2.99, 0.0029)** | (-0.53, 0.596) | **(-3.62, 0.0003)** | (-0.23, 0.822) | (-1.52, 0.128) | **(1.85, 0.064)** | **(-3.37, <0.001)** |
| *LABOR_PCT* | 0.174 | **1.526 ***** | **1.204 **** | -0.738 | **2.692 ***** | **-0.640 *** | **-2.022 ***** | **-0.821 *** |
| | (0.37, 0.709) | **(5.39, <0.001)** | **(2.92, 0.0036)** | (-1.50, 0.133) | **(5.63, <0.001)** | **(-1.98, 0.0476)** | **(-5.16, <0.001)** | **(-2.17, 0.030)** |
| *HH_INCOME* | -0.075 | **-0.237.** | **-0.582 ***** | **0.740 ***** | **-0.344 *** | **0.246 ·** | **-0.430 **** | **0.707 ***** |
| | (-0.37, 0.714) | **(-1.75, 0.080)** | **(-3.71, 0.0002)** | **(3.52, 0.0005)** | **(-2.32, 0.021)** | **(1.85, 0.0645)** | **(-3.01, 0.0027)** | **(3.87, <0.001)** |
| *LU_RESD* | 0.013 | **0.338 ***** | 0.113 | **0.188 ·** | 0.200 | **0.414 ***** | -0.079 | 0.038 |
| | (0.08, 0.934) | **(4.56, <0.001)** | (0.63, 0.529) | **(1.69, 0.092)** | (1.32, 0.187) | **(5.30, <0.001)** | (-0.66, 0.511) | (0.31, 0.753) |
| *LU_RETL* | -0.342 | **-0.368 ***** | 0.057 | -0.035 | **0.229 **** | 0.181 | **-0.339 **** | **0.269 **** |
| | (-1.46, 0.145) | **(-4.77, <0.001)** | (0.45, 0.654) | (-0.40, 0.690) | **(3.01, 0.0027)** | (1.58, 0.115) | **(-2.67, 0.0077)** | **(2.76, 0.0058)** |
| *LU_RECR* | **-0.105 *** | **-0.117** | 0.010 | **-0.205 **** | **0.114 *** | -0.110 | **0.120 ·** | **-0.335 ***** |
| | **(-1.97, 0.049)** | **(-1.05, 0.296)** | (0.12, 0.906) | **(-2.99, 0.0028)** | **(2.15, 0.0319)** | (-0.90, 0.367) | **(1.89, 0.0595)** | **(-9.27, <0.001)** |
| *TRIP_LEN* | **-0.079 ***** | **0.330 ***** | **-0.069 ***** | **0.427 ***** | **-0.052 **** | **0.323 ***** | **-0.078 ***** | **0.364 ***** |
| | **(-4.37, <0.001)** | **(5.55, <0.001)** | **(-4.74, <0.001)** | **(7.13, <0.001)** | **(-3.11, 0.002)** | **(5.62, <0.001)** | **(-5.80, <0.001)** | **(9.34, <0.001)** |
| *NO. OBS* | 1272/59 | 1016/59 | 1272/59 | 1016/59 | 1250/58 | 1004/58 | 1270/59 | 1013/59 |
| *RMSE* | 0.0650 | 0.0712 | 0.0587 | 0.0781 | 0.0572 | 0.0722 | 0.0706 | 0.0814 |
| *MAE* | 0.0454 | 0.0527 | 0.0431 | 0.0562 | 0.0407 | 0.0543 | 0.0536 | 0.0603 |
| *CORR_OBS* | 0.2378 | 0.3775 | 0.3114 | 0.4448 | 0.2526 | 0.4453 | 0.2575 | 0.3847 |
| *LOGLIK* | 1671.46 | 1243.24 | 1802.04 | 1149.09 | 1803.01 | 1214.66 | 1564.94 | 1104.18 |
| *AIC* | -3160.93 | -2312.48 | -3422.08 | -2124.17 | -3426.02 | -2257.32 | -2947.87 | -2034.35 |
| *BIC* | -2692.43 | -1884.12 | -2953.58 | -1695.82 | -2964.24 | -1834.91 | -2479.52 | -1606.25 |

Significance level: ‘***’< 0.001, ‘**’ < 0.01, ‘*’ < 0.05, ‘·’ < 0.1

### *5.2.1 Model Performance*

The 2FE specification captured the large and stable differences across routes and months. On the raw scale, reintroducing route and month means yields very high correlations (above 0.9, not reported in table), confirming that fixed effects soak up most between-route and month-to-month variation. In a fixed-effects context, the more substantive and informative question is how well the covariates explain *within-route*, month-to-month movements after accounting for these fixed differences. As shown in **Table 3**, on this within scale, model performance is modest but meaningful. The observed-fitted correlations range from 0.24 to 0.45, which is typical in noisy panel data settings. For error magnitudes, the model gives within RMSE between 0.057 to 0.081 and MAE between 0.057 to 0.081, indicating relatively small average deviations.

Comparing the model performance between pre-COVID and recovery periods, we see the error magnitudes increased during the recovery period across all time-of-day periods. RMSE rising from 0.065 to 0.071 in the AM-peak, 0.059 to 0.078 in the PM-peak, 0.057 to 0.072 at Midday, and 0.071 to 0.081 in Early evening; the likelihood-based diagnostics also show a decline in log-likelihood and an increase in AIC and BIC values. However, the within observed-fitted correlations increased post-COVID (AM-peak: 0.24 to 0.38; PM-peak: 0.31 to 0.45; Midday: 0.25 to 0.45; Early Evening: 0.26 to 0.38). This observation suggests that, while the variation in ridership became noisier, the covariates used in the models explained the within-route dynamics more effectively during the recovery period. We also see the recovery period with a broader set of significant predictors than the pre-COVID period, indicating that ridership became more sensitive to structural predictors such as service performance variables.

### *5.2.2 Parameter Estimates*

*AM-peak Model*

Before the pandemic, headway was significantly and negatively associated with AM-peak ridership ($\beta = -0.100, p < 0.001$), confirming that service frequency was a key ridership determinant. By contrast, on-time rate was insignificant, which may reflect that many AM-peak riders are core commuters with limited alternatives, relying on transit regardless of its service reliability (Better Streets Miami Beach 2023; Zhao et al. 2014). Among the controls, the share of recreational land use in the route catchment area was negatively related to ridership ($\beta = -0.105, p = 0.049$), consistent with the commute-dominant trip purpose in this period. Notably, trip length showed a negative coefficient ($\beta = -0.079, p < 0.001$) in the pre-COVID period. This may reflect that major route revisions implemented at that time were primarily designed to improve efficiency not merely to reduce coverage (Miami-Dade County 2019). As such, reductions in route length likely signaled more efficient layouts rather than a loss of riders.

In the recovery period, headway was still a significant variable and showed a larger effect ($\beta = -0.202, p = 0.040$). Also, on-time rate emerged as a significant positive ridership determinant ($\beta = 0.877, p = 0.028$). The interaction between headway and on-time rate was also now significant and negative ($\beta = -0.266, p = 0.027$), suggesting that the marginal benefit of punctuality was greater at frequent services with shorter headways. While counter to the common expectation that reliability matters more on infrequent services (Walker 2012), this finding may reflect that, for riders in the study area, frequent services are more sensitive to service punctuality because even small deviations can disrupt tightly scheduled operations and lead to bus bunching or delays. In the recovery period, controls also became more influential with more variables being significant: the white population share ($\beta = -0.204, p = 0.003$) and household income ($\beta =$ $-0.2374, p = 0.08$) of the route catchment area were now negatively associated with ridership, while labor force share ($\beta = 1.526, p < 0.001$) and residential land use share ($\beta = 0.338, p < 0.001$) was positively associated with ridership. Notably, in the recovery period, route trip length remains significant but with a positive sign ($\beta = 0.33, p < 0.001$), suggesting that longer routes were associated with larger AM-peak ridership.

*PM-peak Model*
The PM-peak results closely parallel the AM-peak results. Before COVID, headway was significantly negative ($\beta = -0.11, p = 0.006$), while on-time rate was insignificant. In the recovery period, headway ($\beta = -0.274, p < 0.001$), still significant, showed a larger effect; on-time rate ($\beta = 0.417, p = 0.031$) was now also significant, with a significant negative interaction ($\beta = -0.142, p = 0.0085$) with headway. Compared with AM-peak, the frequency elasticity estimated for PM-peak was larger, while the punctuality elasticity was smaller, reflecting post-work commuters' greater sensitivity to service frequency and smaller sensitivity to punctuality.

Controls showed intuitive patterns. In the pre-COVID period: labor force share was positive ($\beta = 1.204, p = 0.004$), income was negative ($\beta = -0.582, p < 0.001$), and trip length was negative ($\beta = -0.069, p < 0.001$). In the recovery period, the white population share was negative ($\beta = -0.336, p < 0.001$), recreation land use share was negative ($\beta = -0.205, p = 0.003$), and trip length was positive ($\beta = 0.427, p > 0.001$). Interestingly, in the recovery period, higher route-level household income was positively associated with PM-peak ridership ($\beta = 0.740, p < 0.001$). This could be that COVID-19 had disproportionately reduced PM-peak ridership on higher-income routes, while lower-income routes retained a more stable ridership base throughout the pandemic. As ridership gradually returned, the rebound was therefore more visible on higher-income routes, producing the observed positive association in the recovery period.

*Midday Model*
The midday model further reinforced the importance of service performance. Consistent with the two peak-period models, headway was significant in both the pre-COVID period ($\beta = -0.142, p = 0.031$) and the recovery period ($\beta = -0.315, p < 0.001$), and on-time rate was significant in the recovery period ($\beta = 0.584, p = 0.045$) with a negative interaction with headway ($\beta = -0.191, p = 0.006$). Compared with the two peak-period models, the headway coefficients showed larger magnitudes in the midday models, reflecting the relatively less frequent service in this period. The on-time rate coefficient was larger than in the PM-peak model but smaller than in the AM-peak model, indicating that off-peak riders may value service OTP more, though not as much as AM-peak riders. Noteworthy, the variation in on-time rate became significant and positive ($\beta = 0.104, p = 0.071$), indicating that service with inconsistent OTP saw larger ridership. Since variation in OTP often reflects fluctuating traffic conditions (Sun and Chen 2024), the observed positive association may indicate a positive connection between activity demand and transit recovery. This is consistent with the existing observations that travel demand along key corridors rebounded quickly after the pandemic (American Public Transportation Association 2021; Department of Transportation and Public Works, Miami-Dade County 2023).

Control variables revealed distinct patterns from the two peak-period models. In the pre-COVID model for midday period, retail ($\beta = 0.229, p = 0.0027$) and recreational ($\beta = 0.114, p = 0.0319$) land uses were positive, aligning with the midday prevalence of discretionary and shopping trips rather than work-based travel. In the recovery period, labor population share even turned negative ($\beta = -0.640, p = 0.048$). Coupled with residential land use being positive ($\beta = 0.414, p < 0.001$), this finding may indicate shifts toward neighborhood-based travel during the midday period in the COVID recovery period. Trip length and income showed consistent patterns as in the PM-peak model.

*Early Evening Model*

The early evening period diverged from the other three daytime periods. Pre-COVID, all three service performance variables were significant: headway ($\beta = -0.192, p = 0.002$), on-time rate ($\beta = 0.396, p = 0.0257$), and their interaction ($\beta = -0.112, p = 0.0166$). In the recovery period, however, none of the service performance variables were significant ($p > 0.75$). Though, the variation in on-time rate was significant and negative ($\beta = -0.061, p = 0.021$), indicating that consistent service delivery was more important than average punctuality in the evening. This reflects the recent findings on post-COVID travel pattern that riders traveling outside traditional peak hours, especially for non-work purposes, may be more selective and quality-sensitive, thus placing greater value on reliability (Ashour et al. 2024).

The results for controls were more mixed. As expected, labor force share was negative in both before ($\beta = -2.022, p < 0.001$) and after ($\beta = -0.821, p = 0.030$) COVID models, consistent with fewer work trips at night. Trip length and income also showed same patterns as in the PM model. However, the share of white population, retail land use, and recreational land use, showed less consistent associations, positively associated with ridership in some periods but negative in others. These inconsistent patterns suggest that early evening ridership may be influenced by a more heterogeneous set of factors, potentially reflecting a mix of discretionary, social, and work trips, and that contextual and demographic effects are more variable in this period than during daytime hours.

## 6. DISCUSSION

This study examined the associations between route-level ridership and service performance in Miami-Dade County. We focused on two key service performance measures: frequency, operationalized as headway, and punctuality, operationalized as on-time rate and modeled the relationships for the pre-COVID period and recovery period. Our research confirms some findings from the existing literature, underscoring the role of transit service frequency and OTP in shaping ridership. Our analysis further reveals the relative importance of the ridership determinants in the post-pandemic era, providing valuable insights to help transit agencies improve service planning and operations.

### 6.1 Enhanced Role of Service Performance in Shaping Ridership in Recovery Period

We observed a correlation between bus ridership and service performance measures, and this correlation is more significant in the pandemic recovery period compared to the pre-COVID period. Before the pandemic, headway consistently showed a negative association with ridership across the models for all time-of-day periods, while on-time rate was not significant in any. This indicates that frequency was the dominant ridership driver. During the recovery period, for all time-of-day periods except for early evening, headway remained significant, while on-time rate also gained significance. Additionally, the interaction term of the two service performance measures also became significant and positive, suggesting that punctuality has a stronger influence on high-frequency routes. For the early evening period, though headway or on-time rate being insignificant variables, the within-period variation of on-time rate was significant and positive, indicating greater consistency in OTP was associated with larger ridership.

The finding that service frequency remained a critical determinant of ridership across both periods, with even larger effects during the recovery period, reaffirms the recent findings that riders are sensitive to service frequency and that service reductions could trigger a "doom spiral" of declining ridership in the post-pandemic era (Redelmeier and El-Geneidy 2024). Our research provides empirical evidence supporting federal and local subsidies to maintain and improve transit service frequencies in the post-pandemic era, aiding in ridership growth.

The newly observed significant association with on-time rate, on the other hand, stressed the importance of punctual service. This is consistent with survey findings that underscore riders' stronger preference for timely transit service after the pandemic (Hsieh 2023; Redelmeier and El-Geneidy 2024). To translate these insights into actionable strategies, transit agencies can consider a comprehensive approach that integrates technology, infrastructure, and operational policies. First, the implementation of technology-driven real-time bus information systems can help reduce rider anxiety and reinforce perceptions of service reliability. This can be achieved by deploying live schedule displays at bus stops or creating a unified mobile application that provides real-time updates on schedules, expected arrivals and delays (Ashour et al. 2024). Second, in terms of infrastructure, transit agencies can prioritize the expansion of Bus Rapid Transit (BRT) corridors, signal-priority systems, and dedicated bus lanes, especially in ridership-dense areas identified in this study. These transit-priority measures are essential for mitigating delays caused by traffic congestion and enhancing overall service reliability (Alex Roman 2024; Christopher 2022; Subbarao and Kadali 2022). Third, operational strategies must also be refined to ensure immediate and effective responses to service delays. Dynamic crew allocation policies can be codified to ensure drivers receive timely feedback and support when they are at risk of falling behind schedule (Huisman 2004). Additionally, implementing incentives such as a 5-minute "service-recovery" alert can mobilize driver teams to proactively address delays, thereby maintaining headway and enhancing service reliability (S.Dukare et al. 2015).

Taken together, these service performance improvement measurements not only offer promising solutions to increase service reliability and boost both passenger satisfaction and ridership, but they also embed a resilient service paradigm ready for future demand shocks. (Subbarao and Kadali 2022).

### 6.2 Relationship between Ridership and Service Performance Vary by Time of Day

In this study, we estimated the log-log models separately for each time-of-day period, using period-specific variables to obtain elasticities for each period. This enables us to compare the coefficients across periods, generating insights into whether service characteristics have stronger or weaker associations with ridership by time of day. The three daytime periods (AM-peak, PM-peak, and Midday) displayed consistent patterns as discussed earlier: headway was always a significant negative determinant of ridership, both before and after COVID, while on-time rate was insignificant in the pre-COVID period but became significant in the recovery period. Comparing magnitudes, however, we observed differences between time-of-day periods. The on-time rate coefficient was the highest during the AM-peak, reflecting the time-sensitive nature of work-related trips, whereas its magnitude was smallest in the PM-peak, likely due to the mixed purposes of evening travel. Headway elasticity was the greatest in the Midday period, where lower service levels make frequency changes more impactful, and smallest in the AM-peak, suggesting that frequency outweighs OTP concerns for Midday riders. The PM-peak exhibited intermediate sensitivity to frequency, higher than the AM-peak but lower than midday, highlighting the importance of maintaining steady service to support diverse travel needs.

These findings highlight the need for transit agencies to consider targeted ridership-boosting strategies that differ by time of day. For AM-peak, given that on-time rate is a highly significant performance measurement, agencies can therefore focus on reliability-enhancing tactics such as signal priority, bus-only lanes. Agencies may also explore offering express or limited-stop variants to cut travel time and rapid crew response to keep buses on schedule. In midday, given that the elasticity of frequency is large, it is important to maintain a minimum level of frequency. Agencies need to find the proper pulse of the service, maintain adherence to the schedule, and effectively communicate this information to riders (Walker 2012). For PM-peak, while OTP has less influence on ridership compared to the AM-peak and Midday, it remains a significant factor, and frequency continues to be a key determinant. Previous post-COVID studies show that PM-peak travel now encompasses a broader range of trip purposes beyond commuting, and key corridors already continue to experience significant congestion (Speroni et al. 2024). Strategies for PM-peak operation should therefore focus on maintaining sufficient service levels on core corridors to meet PM-peak demand, and, at the same time, adjusting schedules to accommodate these diverse travel needs. In summary, by weaving these period-specific strategies, transit agencies can strengthen overall service performance and more effectively nurture ridership in a post-pandemic era and beyond.

## 7. CONCLUSIONS

Our study analyzed bus ridership trends in Miami-Dade County and modeled its relationship with service frequency and on-time performance (OTP). Utilizing Automatic Passenger Count (APC) data alongside stop arrival records from General Transit Feed Specification Real Time (GTFS-RT) data, we constructed a panel dataset and developed two-way fixed-effects (2FE) models with headway and on-time rate as key explanatory variables. We estimated separate models for four time-of-day periods: AM-peak, Midday, PM-peak, and Early evening and in both a pre-COVID period and a post-COVID recovery period.

Our findings underscore the critical role of service frequency and the increased importance of OTP in shaping ridership, particularly in the post-pandemic era. The results reveal that while frequency remained a critical determinant of ridership in both periods, its effect was more pronounced during the recovery period. Additionally, the significance of punctuality in this period suggests that ridership is more responsive to service OTP. These insights have practical implications for transit planning, advocating for the maintenance and enhancement of transit service reliability, especially through measures like Bus Rapid Transit, signal prioritization, and dedicated bus lanes.

The study also highlights temporal heterogeneity in the relationship between service performance and ridership, with varying elasticities across different times of day. Notably, midday periods exhibited the largest frequency elasticity, while the AM-peak period showed the highest sensitivity to on-time rate. These findings suggest the necessity for transit agencies to tailor service to the specific time-of-day patterns: targeting OTP for morning commutes, frequency for midday travel and PM-peak. These strategies may effectively boost ridership, cater to diverse travel behaviors, and enhance system resilience.

Overall, our analysis illuminates the evolving ridership dynamics in the post-COVID landscape. The increased variability in ridership and the heightened sensitivity to service performance call for a reevaluation of transit usage pattern and planning approaches. Emphasizing flexibility and adaptability will be essential in addressing the more complex and varied demands of modern transit users. As transit agencies navigate this new terrain, our study provides valuable guidance on how targeted operational improvements can support ridership recovery and promote a more robust and responsive transit system in the future.

This study has several limitations. Although we included sociodemographic and land use controls, the analysis is conducted at an aggregated route level, which may mask finer-grained variations. Additionally, changes in in rider behavior and travel patterns like telecommuting are difficult to observe at route level; finer-grained stop- or OD-level analyses may provide deeper insights. It is also important to note that insignificant results do not imply these variables are unimportant; they may instead reflect the limitations of the models. The 2FE model assumes homogeneity in the impact of service performance measures on ridership across routes, potentially oversimplifying the complex relationships across different routes. Larger datasets could also enable the use of more flexible modeling approaches, such as mixed-effects models, which allow for heterogeneous responses across routes or riders (Redelmeier and El-Geneidy 2024; Yaffee 2003). Future research may also complement the operational data used here with individual-level data (e.g., periodic transit user surveys) and carry out analysis to gain deeper insights into the shifts in rider characteristics, travel needs, and personal choice factors. Such data can verify and extend our study findings, enabling transit authorities to respond more effectively to the evolving rider needs.

## References

Aemmer, Z., Ranjbari, A., MacKenzie, D.: Measurement and classification of transit delays using GTFS-RT data. Public Transp. 14, 263–285 (2022). https://doi.org/10.1007/s12469-022-00291-7

Alex Roman: Public Transit Ridership Continues Post-COVID Bounce Back, https://www.metro-magazine.com/10216134/public-transit-ridership-continues-post-covid-bounce-back, (2024)

American Public Transportation Association: On the Horizon: Planning for Post-Pandemic Travel. (2021)

Arhin, S., Noel, E.C.: Evaluation of bus transit reliability in the District of Columbia. Mineta National Transit Research Consortium (2013)

Ashour, L.A., Shen, Q., Moudon, A., Cai, M., Wang, Y., Brown, M.: Post-pandemic transit commute: Lessons from focus group discussions on the experience of essential workers during COVID-19. J. Transp. Geogr. 116, 103832 (2024). https://doi.org/10.1016/j.jtrangeo.2024.103832

Berrebi, S.J., Joshi, S., Watkins, K.E.: On bus ridership and frequency. Transp. Res. Part Policy Pract. 148, 140–154 (2021). https://doi.org/10.1016/j.tra.2021.03.005

Better Streets Miami Beach: Better Streets Miami Beach, https://www.change.org/p/build-the-baylink-metromover-to-south-beach, (2023)

Bowman, L.A., Turnquist, M.A.: Service frequency, schedule reliability and passenger wait times at transit stops. Transp. Res. Part Gen. 15, 465–471 (1981). https://doi.org/10.1016/0191-2607(81)90114-X

Boyle: Fixed-Route Transit Ridership Forecasting and Service Planning Methods. Transportation Research Board, Washington, D.C. (2006)

CBS Miami: New SR 836 Express Metro Bus Service To Provide Rush Hour Relief, https://www.cbsnews.com/miami/news/sr836-express-metro-bus-service-rush-hour/, (2020)(a)

CBS Miami: Coronavirus Impact: Miami-Dade Transit Suspending Bus Routes, https://www.cbsnews.com/miami/news/coronavirus-miami-dade-transit-suspending-bus-routes/, (2020)(b)

Chen, C., Varley, D., Chen, J.: What Affects Transit Ridership? A Dynamic Analysis involving Multiple Factors, Lags and Asymmetric Behaviour. Urban Stud. 48, 1893–1908 (2011). https://doi.org/10.1177/0042098010379280

Christopher, C.: How US cities are navigating post-COVID transit challenges, https://cities-today.com/how-us-cities-are-navigating-post-covid-transit-challenges/, (2022)

County of Miami-Dade Transportation & Public Work: 2021-07-28 Existing Bus Network Peak, https://www.miamidade.gov/transportation/library/existing-bus-network-peak-routes.pdf, (2021)

Croissant, Y., Millo, G.: Panel data econometrics in R: The plm package. J. Stat. Softw. 27, 1–43 (2008)

Danaher, A.: Transit Capacity and Quality of Service Manual—3rd Edition. Transportation Research Board (2003)

Dasmalchi, E., Taylor, B.D.: Examining Shifts in the Balance of Riders and Bus Service Before and During the Pandemic in Boston, Houston, and Los Angeles. Findings. (2022). https://doi.org/10.32866/001c.34216

Department of Transportation and Public Works, Miami-Dade County: Annual Report 2023. (2023)

Diab, E., DeWeese, J., Chaloux, N., El-Geneidy, A.: Adjusting the service? Understanding the factors affecting bus ridership over time at the route level in Montréal, Canada. Transportation. 48, 2765–2786 (2021). https://doi.org/10.1007/s11116-020-10147-3
Erhardt, G.D., Hoque, J.M., Goyal, V., Berrebi, S., Brakewood, C., Watkins, K.E.: Why has public transit ridership declined in the United States? Transp. Res. Part Policy Pract. 161, 68–87 (2022). https://doi.org/10.1016/j.tra.2022.04.006
Espinoza, L.: Miami-Dade Transit increases services as Miami-Dade County moves to a New Normal, https://www.miamidade.gov/releases/2020-05-15-dtpw-reopen-service-adjustments.asp, (2020)
Genna, G.: Swiftly Help Center - API Guide Basics, https://swiftly.zendesk.com/hc/en-us/articles/360049238811-API-Guide-Basics
Gkiotsalitis, K., Cats, O.: Public transport planning adaption under the COVID-19 pandemic crisis: literature review of research needs and directions. Transp. Rev. 41, 374–392 (2021). https://doi.org/10.1080/01441647.2020.1857886
Hall, J.D., Palsson, C., Price, J.: Is Uber a substitute or complement for public transit? J. Urban Econ. 108, 36–50 (2018). https://doi.org/10.1016/j.jue.2018.09.003
Hsieh, H.-S.: Understanding post-COVID-19 hierarchy of public transit needs: Exploring relationship between service attributes, satisfaction, and loyalty. J. Transp. Health. 32, 101656 (2023). https://doi.org/10.1016/j.jth.2023.101656
Hu, S., Chen, P.: Who left riding transit? Examining socioeconomic disparities in the impact of COVID-19 on ridership. Transp. Res. Part Transp. Environ. 90, 102654 (2021). https://doi.org/10.1016/j.trd.2020.102654
Huisman, D.: Integrated and dynamic vehicle and crew scheduling. (2004)
Kapatsila, B., Bahamonde-Birke, F.J., Van Lierop, D., Grisé, E.: Impact of the COVID-19 pandemic on the comfort of riding a crowded bus in Metro Vancouver, Canada. Transp. Policy. 141, 83–96 (2023). https://doi.org/10.1016/j.tranpol.2023.07.018
Kerkman, K., Martens, K., Meurs, H.: Factors Influencing Stop-Level Transit Ridership in Arnhem–Nijmegen City Region, Netherlands. Transp. Res. Rec. J. Transp. Res. Board. 2537, 23–32 (2015). https://doi.org/10.3141/2537-03
Kittelson & Associates, Inc., Brinckerhoff, P., KFH Group, Inc., Texas A&M Transportation Institute, Arup, Transit Cooperative Research Program, Transportation Research Board, National Academies of Sciences, Engineering, and Medicine: Transit Capacity and Quality of Service Manual, Third Edition. Transportation Research Board, Washington, D.C. (2013)
Lee, J., Porr, A., Miller, H.: Evidence of Increased Vehicle Speeding in Ohio's Major Cities during the COVID-19 Pandemic. Findings. (2020). https://doi.org/10.32866/001c.12988
Liu, L., Miller, H.J., Scheff, J.: The impacts of COVID-19 pandemic on public transit demand in the United States. PLOS ONE. 15, e0242476 (2020). https://doi.org/10.1371/journal.pone.0242476
Liu, L., Porr, A., Miller, H.J.: Measuring the impacts of disruptions on public transit accessibility and reliability. J. Transp. Geogr. 114, 103769 (2024). https://doi.org/10.1016/j.jtrangeo.2023.103769
Lucas Albuquerque-Oliveira, J., Moraes Oliveira-Neto, F., Pereira, R.H.M.: A novel route-based accessibility measure and its association with transit ridership. Transp. Res. Part Policy Pract. 179, 103916 (2024). https://doi.org/10.1016/j.tra.2023.103916
Lucas, K.: Transport and social exclusion: Where are we now? Transp. Policy. 20, 105–113 (2012). https://doi.org/10.1016/j.tranpol.2012.01.013

Miami-Dade County: Report on October 2018 Cycle Applications to Amend the Comprehensive Development Master Plan, https://www.miamidade.gov/govaction/legistarfiles/Matters/Y2019/190856.pdf, (2019)
Preacher, K.J., Rucker, D.: A primer on interaction effects in multiple linear regression, (2003)
Redelmeier, P., El-Geneidy, A.: If You Cut It Will They Ride? Longitudinal Examination of the Elasticity of Public Transport Ridership in the Post-Pandemic Era. Transp. Res. Rec. J. Transp. Res. Board. 03611981241240754 (2024). https://doi.org/10.1177/03611981241240754
S.Dukare, S., A. Patil, D., P. Rane, K.: Vehicle Tracking, Monitoring and Alerting System: A Review. Int. J. Comput. Appl. 119, 39–44 (2015). https://doi.org/10.5120/21107-3835
Shimamoto, H., Kusubaru, R.: Evaluation of the short- and long-term impacts of the COVID-19 pandemic on bus ridership in Miyazaki City, Japan. Asian Transp. Stud. 9, 100098 (2023). https://doi.org/10.1016/j.eastsj.2023.100098
Shortall, R., Mouter, N., Van Wee, B.: COVID-19 passenger transport measures and their impacts. Transp. Rev. 42, 441–466 (2022). https://doi.org/10.1080/01441647.2021.1976307
Soza-Parra, J., Muñoz, J.C., Raveau, S.: Factors that affect the evolution of headway variability along an urban bus service. Transp. B Transp. Dyn. 9, 479–490 (2021). https://doi.org/10.1080/21680566.2021.1906350
Speroni, S., Siddiq, F., Paul, J., Taylor, B.D.: Peaked too soon? Analyzing the shifting patterns of PM peak period travel in Southern California. Travel Behav. Soc. 36, 100787 (2024). https://doi.org/10.1016/j.tbs.2024.100787
Subbarao, S.S.V., Kadali, R.: Impact of COVID-19 pandemic lockdown on the public transportation system and strategic plans to improve PT ridership: a review. Innov. Infrastruct. Solut. 7, 97 (2022). https://doi.org/10.1007/s41062-021-00693-9
Sun, Y., Chen, Y.: Travel Time Variability in Urban Mobility: Exploring Transportation System Reliability Performance Using Ridesharing Data. Sustainability. 16, 8103 (2024). https://doi.org/10.3390/su16188103
Taylor, B.D., Fink, C.N.: The factors influencing transit ridership: A review and analysis of the ridership literature. (2003)
TransitCenter: Your Bus Is On Time. What Does That Even Mean?, https://transitcenter.org/bus-time-even-mean/, (2018)
University of Florida GeoPlan Center: Generalized Land Use Derived From 2023 Florida Parcels, https://fgdl.org/zips/metadata/xml/lu_gen_2023.xml, (2024)
U.S. Census Bureau: American Community Survey 5-Year Estimates (2018–2022), Miami-Dade County, Florida, https://data.census.gov/table?q=Miami-Dade%20County,%20Florida&d=ACS%205-Year%20Estimates%20Data%20Profiles, (2022)
U.S. National Transit Database: Complete Monthly Ridership (with adjustments and estimates), https://www.transit.dot.gov/ntd/data-product/monthly-module-raw-data-release, (2024)
Walker, J.: Human Transit. Island Press/Center for Resource Economics, Washington, DC (2012)
Warade, R.: How Data Quality Impacts Service Reliability and the Passenger Experience, https://www.goswift.ly/blog/transforming-transit-with-high-quality-data, (2025)
Wilbur, M., Ayman, A., Sivagnanam, A., Ouyang, A., Poon, V., Kabir, R., Vadali, A., Pugliese, P., Freudberg, D., Laszka, A., Dubey, A.: Impact of COVID-19 on Public Transit Accessibility and Ridership. Transp. Res. Rec. J. Transp. Res. Board. 2677, 531–546 (2023). https://doi.org/10.1177/03611981231160531
Wooldridge, J.M.: Econometric analysis of cross section and panel data. MIT press (2010)

Yaffee, R.: A primer for panel data analysis. Connect Inf. Technol. NYU. 8, 1–11 (2003)
Yu, C., Dong, W., Liu, Y., Yang, C., Yuan, Q.: Rethinking bus ridership dynamics: Examining nonlinear effects of determinants on bus ridership changes using city-level panel data from 2010 to 2019. Transp. Policy. 151, 85–100 (2024). https://doi.org/10.1016/j.tranpol.2024.04.004
Zhao, J., Webb, V., Shah, P.: Customer Loyalty Differences between Captive and Choice Transit Riders. Transp. Res. Rec. J. Transp. Res. Board. 2415, 80–88 (2014). https://doi.org/10.3141/2415-09

**Vitae/Biography**

**Duanya Lyu** is a PhD student in the Department of Civil and Coastal Engineering at the University of Florida, Florida, U.S. Her work centers on transit, micromobility, and human mobility, using data-driven analysis. She is working toward evidence-based insights that can support transportation policy.

**Xiang (Jacob) Yan** is an Assistant Professor in the Department of Civil and Coastal Engineering at the University of Florida, Florida, U.S. His work focuses on leveraging data-driven and AI-enabled methods to advance public transit and sustainable transportation systems. Bridging urban planning, data science, and transportation engineering, he takes an interdisciplinary approach to transportation research that connects technological innovations with community needs.